\documentclass[10pt,twocolumn,letterpaper]{article}

\usepackage{wacv}
\usepackage{times}
\usepackage{epsfig}
\usepackage{graphicx}
\usepackage{amsmath}
\usepackage{amssymb}
\usepackage{float}
\usepackage{subfigure}
\usepackage{adjustbox}

\wacvfinalcopy 

\ifwacvfinal
\def\assignedStartPage{1} 
\fi

\ifwacvfinal
\usepackage[breaklinks=true,bookmarks=false]{hyperref}
\else
\usepackage[pagebackref=true,breaklinks=true,colorlinks,bookmarks=false]{hyperref}
\fi

\ifwacvfinal
\else
\fi

\begin{document}

\title{Deep Learning based Dimple Segmentation for Quantitative Fractography}

\author{Ashish Sinha\\
Department of Metallurgical and Materials Engineering\\
Indian Institute of Technology Roorkee\\
Roorkee-247 667, India\\
{\tt\small asinha@mt.iitr.ac.in}
\and
K.S Suresh\\
Department of Metallurgical and Materials Engineering\\
Indian Institute of Technology Roorkee\\
Roorkee-247 667, India\\
{\tt\small ks.suresh@mt.iitr.ac.in}
}

\maketitle

\begin{abstract}
   In this work, we try to address the challenging problem of dimple segmentation in titanium alloys using machine learning methods, especially neural networks. The  fractographic images for this task are obtained using a Scanning Election Microscope (SEM). To determine the cause of fracture in metals we address the problem of segmentation of dimples in fractographs i.e. the fracture surface of metals using supervised machine learning methods. Determining the cause of fracture would help us in material property, mechanical property prediction and development of new fracture-resistant materials. This method would also help in correlating the topographical features of the fracture surface with the mechanical properties of the material. Our proposed novel model achieves the best performance as compared to other previous approaches. To the best of our knowledge, this is one of the first work in fractography using fully convolutional neural networks with self-attention for supervised learning of deep dimple fractography, though it can be easily extended to account for brittle characteristics as well.
\end{abstract}

\section{Introduction}

Titanium is an important metal for making the plates of body armour of soldiers, body implants, surgical instruments. In addition, titanium alloys are also used for making aircrafts and spacecrafts due to it's high strength and wear resistance \cite{attar2014manufacture} \cite{ehtemam2016evaluation} \cite{kabashkin2010reliability}.

Fracture patterns of metals in general and high-strength titanium and iron alloys in particular happens in a stage-like nature during deformation and stress accumulation. Deformation of metals caused due to application of load or corrosive actions of nature, causes an accumulation of pores predominantly in the central part of the neck of the fracture, which coalesce with grain (can be thought as domains in magnetic field) conglomerates leading to the growth of the crack in a continuous fashion in the direction of loading. Thus, the central crack which grows by thinning and breaking connections between the pores, together with the newly formed crack leaves traces on the surface in the form of dimples, which indicates the history of the material fracture \cite{beachem1973elastic} \cite{kardomateas1986fractographic} \cite{merson2017confocal}.

\begin{figure}[h]
    \centering
    \includegraphics[scale=0.5]{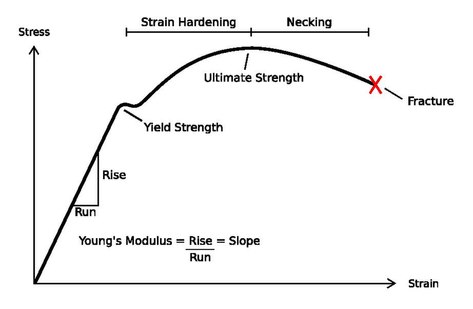}
    \caption{Stress-Strain curve}
    \label{fig:stresscurve}
\end{figure}

The process of damage to a material can be depicted on stress-strain curve, whereas fracture is the final stage of deformation \ref{fig:stresscurve}. These links between the stages of deformation are important when analyzing the causes of fracture using fractographic analysis. Fractographic analysis uses physics of solid body, material science, optic-digital methods to determine the causes of fracture. Earlier, parameter measurements of fracture surfaces were made manually or automatically but the software was positioned by a operator. A large variety of materials made the generalization difficult \cite{bastidas2016fractographic}.

\begin{figure}[h]
    \centering
    \includegraphics[scale=0.5]{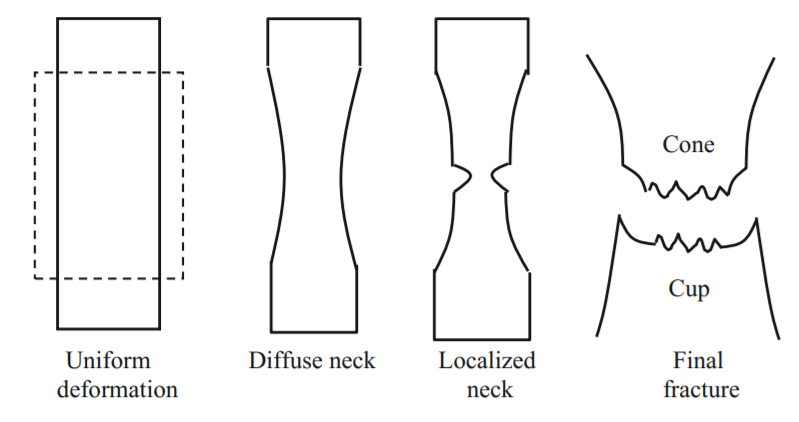}
    \caption{Steps of ductile fracture under stress}
    \label{fig:ductile1}
\end{figure}

Administered by both the extrinsic (e.g. imposed loading, environmental conditions) and the intrinsic (microstructure) characteristics is the process of fracture of materials. The data regarding the affect of both intrinsic and extrinsic characteristics of the fracture process is contained in the surface of the fracture. Important means applied to study the surface of fracture is fractography and obtain factors approaching to failure and data relating properties of material. To attain the topographic characterization of the surface, fractography is used. Analyzing and classifying the  several mechanisms of fracture, and the interrelationship between them with the microstructure of material,the situations approaching towards it's failure, and its mechanical nature are also some applications in which fractography plays its part.

Formation of dimples on the fracture surface is due to the ductile fracture of materials. Generally, the micro-pore merger in the material during deformation in plastic region leads to such dimples. After effect of the pore rupture and the destruction of the surrounding material are dimples as shown in \ref{fig:ductile1}. Many models of nucleation, growth, and coalescence of pores are known and their application under uncertain condition leads to fracture of materials, the results of which are usually complicated to compare.

The ponderous process which is quantitative and qualitative assessment of fracture surface is executed manually by experienced and well-trained technicians, hence requiring significant labour. Quantitative estimations are likely to have more errors because there is a huge dependency on human factor.


\section{Related Work}

In the last century, metallurgists and material scientists have acquired, analysed and compared images of micro-structures in a progressive way. 
Most of the effort was put in to learn a defined material system and different classes of micro structures.if the features of micro structures are known then digital analysis techniques on images would be helpful in characterising, segmenting and comparing different structures with high resolution. Presently the analytics of microstructures is mostly focused on finding the relationship between structure and properties by using the shape and size and appearance of the features. These approaches have moved forward to more of a machine learning approach where the properties are found out by the choosing of correct algorithms.
Some of the areas where machine learning techniques are used in metallurgy and materials science are: new material design, material property prediction, microstructure recognition, and analysis of failure in a material, etc.

\subsection{Computer Vision in Quantitative Fractography}

The main interest in fractography is due in finding the correlation between the 
features of the surface that is fractured and the environment or conditions that lead to its failure. For centuries, this has remained qualitative in nature. Mostly, scientists examine the Scanning Electron Microscope (SEM), Optical Microscope (OPM) images of samples for failure analysis.
Therefore quantitative analysis brings the potential for improving and understanding the mechanisms that control the fracture process and also determine the reliability of the models that in the current material design system. With the latest advancement in the field of image analysis and moreover the availability of machine learning tools, it has become more easy to automate the event of finding important features and information from the fractographs. 

The use of computer vision methodologies by \cite{hu2017automation} \cite{chowdhury2016image} lead to identify the images which contained dendritic morphology to classify if the the direction was longitudinal or transverse. Another use case has been automatic measuring the volume of ferritic(iron) volume fraction from the binary phase structures of ferritic and austenite (a phase of iron). In the field of fractography, many successful attempts have been made to build automated models for quantitative fractography \cite{bastidas2016fractographic} used non-linear algorithms of machine learning (ANN and SVM) and combined it with texture analysis to classify the images into there modes: ductile sudden, brittle sudden and fatigue. A recent work \cite{tsopanidis2020toward} aims to quantify fracture surface for materials with brittle fracture characteristics. Our work focuses primarily on ductile materials.

In our project, we use different variants of deep neural networks to segment deep dimples and benchmark the results, in ductile fracture materials on SEM images of titanium alloys which can be further used to find the properties of fracture mechanism.

In previous works, the authors have proposed methods for fractographic recognition, control and calculation of parameters of the dimples of based on neural networks \cite{konovalenko2017fuzzy}  \cite{maruschak2018impact}. In \cite{konovalenko2018investigation}, the authors trained 17 models of neural networks with various sets of hyper-parameters, then their speed and accuracy were evaluated and the optimal neural network was selected. We propose a fully convolutional U-Net \cite{ronneberger2015u} inspired deep neural network with position and channel based attention residual blocks with dense connections \cite{woo2018cbam} \cite{he2016deep} \cite{huang1608densely} and squeeze and excitation \cite{hu2018squeeze} block in the bottleneck layer for the segmentation of deep dimples.

Our work explores the application of deep learning methods in fractography, an active field of research in material science. Below, we briefly explain the terms necessary to better understand our work.

\subsection{Fractography}

Fractography is a technique to understand the causes of failures and also to verify theoretical failure predictions with real life failures. It can be used in forensics, for analyzing broken products which have been used as weapons, such as broken bottles. Thus, a defendant might claim that a bottle was faulty and broke accidentally when it impacted a victim of an assault. Fractography could show the allegation to be false and that considerable force was needed to smash the bottle before using the broken end as a weapon to deliberately the victim. In these cases, the overall pattern of cracking is important in reconstructing the sequence of events, rather than the specific characteristics of a single crack, since crack grows by coalescing with other grains in the microstructure of the metals. 

\subsection{Crack Growth}

The initiation and continuation of crack growth is dependant on several factors such as bulk material properties, geometry of the body, geometry of the crack, loading rate, loading distribution, load magnitude, environmental conditions, time and microstructure. Cracks are initiated, and as the cracks grow, energy is transmitted to the crack tip at an energy release rate $G$, which is a function of the applied load, crack length and the geometry of the body. All solid materials, have an intrinsic energy release rate $G_C$, where $G_C$ is referred to as the fracture energy or fracture toughness of the material. A crack will grow if {$G$ $\geq$  $G_C$}.

\subsection{Microstructure}

Microstructure is a very small scale structure of a material, defined as the structure of a prepared surface of a material as revealed under an optical microscope above 25x magnification. The microstructure of a material correlates strongly with the strength, toughness, ductility, hardness, wear resistance, etc of the material. A microstructure's influence on the physical and mechanical properties of a material is governed by the different defects present or absent in the structure. These defects can take many forms but the primary ones are the pores. To acquire micrographs, both optical as well as electron microscopy is used. 

\subsection{Dimple Fracture}

A dimple fracture is a type of material failure on a metal's surface that is characterized by the formation and collection of micro-voids along the granular boundary of the metal i.e. the fracture path. The occurrence of dimple fractures  is directly proportional to increased corrosion rates. The material appears physically dimpled when examined under high magnification. There are three main types of dimple fractures: 

\begin{itemize}
    \item Shear fractures 
    \item Tearing fractures 
    \item Tensile fractures
\end{itemize}

All three of these fractures are characterized by tiny holes, known as micro-voids, which are microscopically located in the interior of a piece of metal when under the force of an external load. The greater the load, the greater the proximity and the total gap volume of these voids. The appearance of such a fractured surface is referred to as a dimple rupture. A scanning electron microscope can be used to examine a dimple rupture at a magnification of about 2500x.



\section{Methodology}

Here we discuss the fractographic analysis for detection of dimples in fracture metal surfaces using various deep learning models. We consider Titanium (Ti) alloys as our focus of discussion. In the next section we briefly discuss the previous methods used for segmentation tasks but which are new in this domain, and then we explain our proposed model. 

\subsection{Previous Approaches}

\subsubsection{U-Net}

The U-Net \cite{ronneberger2015u} is mainly employed for bio-medical image segmentation.
It has two parts: a contracting encoder and an expanding decoder. The encoder, a feature correction path, is a continuous stack of convolution and pooling layers; used for image identification.
The decoder, a feature expanding path, is used for collecting exact localisation of fractures using transposed convolutions or deconvolutions. The model is an end-to-end fully convolutional network. The image of any size can be fed in the model as it lacks any densely connected layer.

\subsubsection{U-Net++}

This model uses dense  block ideas of DenseNet \cite{huang1608densely} to improve upon U-Net \cite{ronneberger2015u}.  

    



This model architecture consists of an encoder sub-network and a decoder sub-network after it. The skip connections between each node comprises of three convolution layers and a deep dense  convolution block. A convolution layer follows every concatenation layer. This convolution layer considers output from preceding convolution layers of the dense block and up-sampled output from lower dense block, and fuses them.

 
\subsubsection{Mask R-CNN}

Mask R-CNN \cite{he2017mask} is the skeletal that assists in object detection and localization. It completes the task in two steps:   scanning the image and generating proposals to point the probable locations of an object,\textcolor{white}{b}classification on first step proposals, and generation of bounding boxes and masks.

\subsection{Proposed Model}

\begin{figure*}[t]
\begin{center}
\includegraphics[scale=0.7]{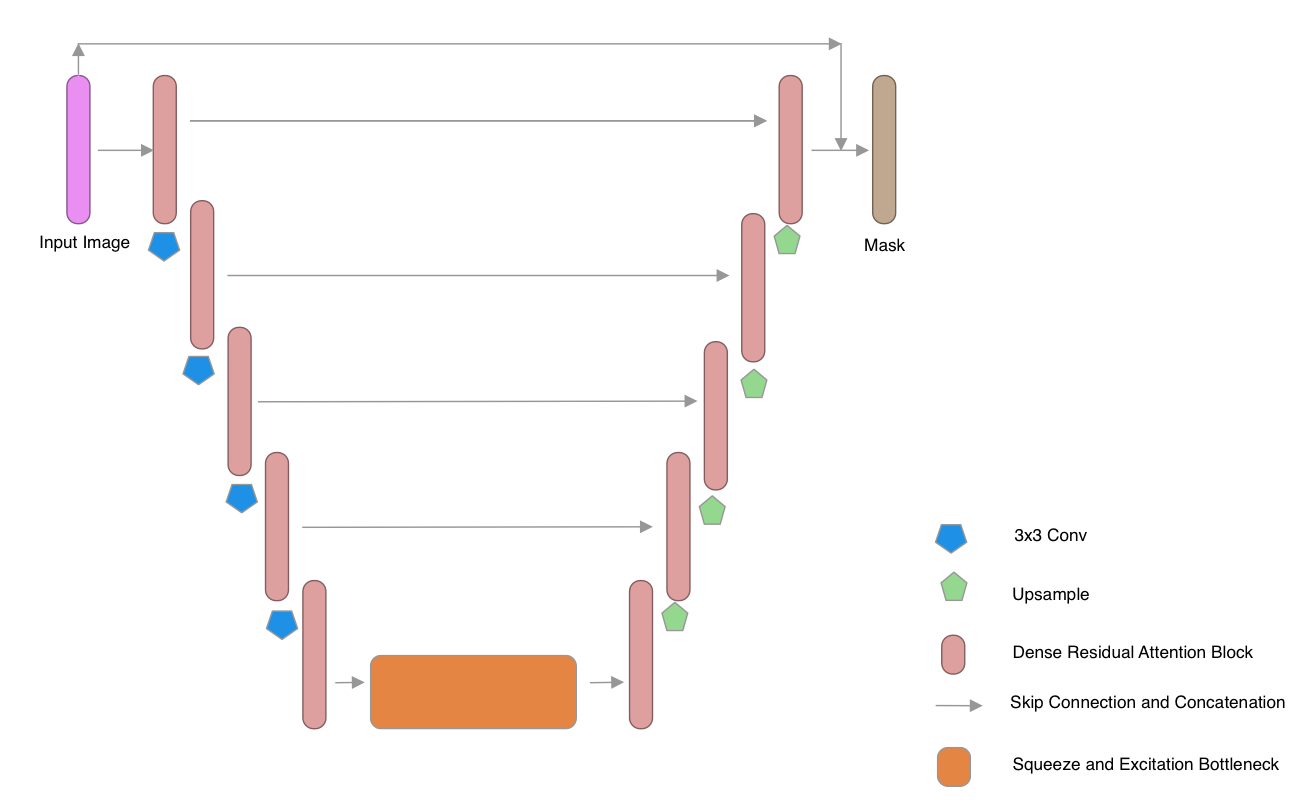}
\end{center}
   \caption{Our Proposed model}
\label{fig:proposed_model}
\end{figure*}

Like the U-Net architecture, our proposed network is divided into two parts, a contracting encoder and an expanding decoder. For the encoder part, we employ 5 layers of residual convolutional blocks with channel and spatial attention as proposed in \cite{fu2019dual} but instead of using attention blocks in parallel we find that using them as proposed in \cite{woo2018cbam} gives better results in our case. We employ a dense connection \cite{huang1608densely} in each residual block used in the encoder of the model. Bottleneck blocks consists of 3 layers of dense connection of residual convolution blocks followed by a squeeze and excitation block \cite{hu2018squeeze}. The encoder part starts with a convolution of filter size 5 and stride 1 followed by a batch normalization layer \cite{ioffe2015batch} and a PRelu activation. The other convolutions have a filter size of 3.

In the decoder part, we make use of the residual convolutional blocks similar to that of the encoder and use parameter-free bilinear upsampling instead of transposed convolutional operations to reduce the number of trainable parameters  \cite{de2018clinically}. The overall model architecture is shown in \ref{fig:proposed_model}. Each upsampling block is followed by a channel and spatial attention block.

The goal of this work is not to propose a novel architecture but to establish a baseline for further development and also to show the application of deep learning models on an age old problem of dimple detection for quantitative fractography so as to determine the cause of fracture in materials which in turn will lead to the design of new and better materials. This work can be a reference for the material scientists to further explore the domain of fractography with machine learning. 

\section{Evaluation}
\subsection{Dataset}

\begin{figure}[H]
    \centering
    \includegraphics[width = 0.45\linewidth]{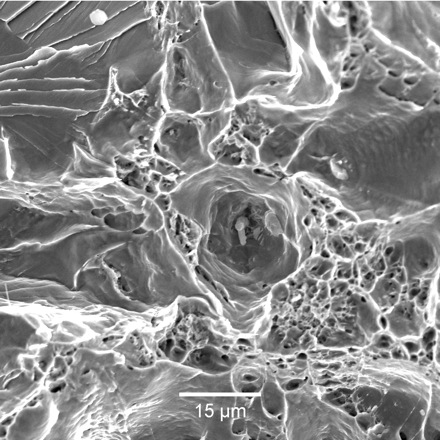}\hfill
    \includegraphics[width = 0.45\linewidth]{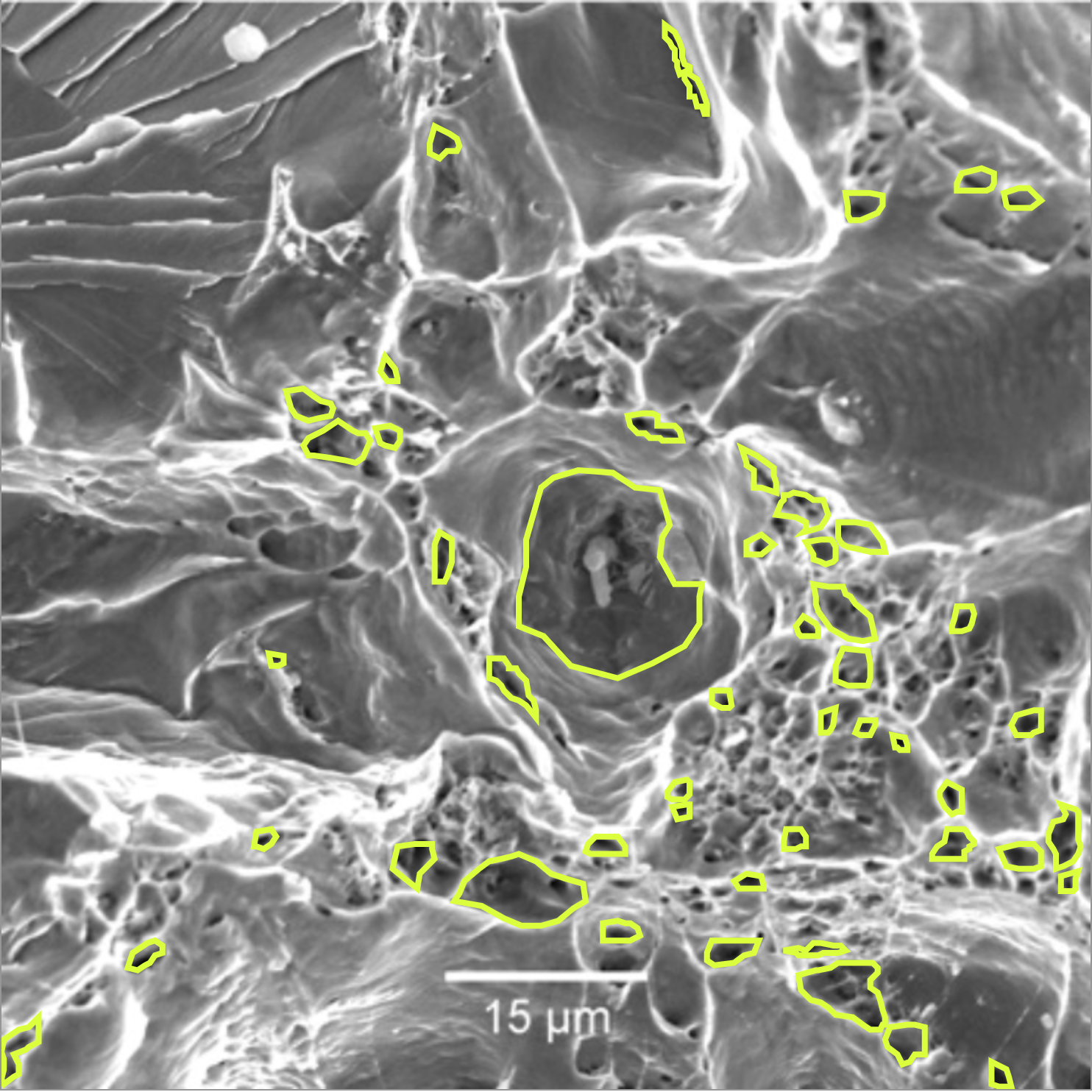}
    \caption{A sample fractograph and it's GT showing dimples}
    \label{fig:my_label}
\end{figure}

For conducting our experiments, we collected 216 high-resolution SEM images of Ti alloys (tested under various physical conditions) at 200x, 500x, 1000x and 2000x magnification. It is a very difficult process to obtain this kind of data in such magnitude since it requires heavy pre-processing with chemicals of the metal surface before it can be viewed under a scanning electron microscope. Typically, it required around a day for the polishing and obtaining a SEM image of the metal surface. We performed the extensive tedious task of annotating the SEM images with deep dimples. During annotation, we classify the areas of the SEM images that presented the most characteristic features of deep dimple (dark areas), while the areas with unclear classification or ambiguous features are labeled as background. As visible from \ref{fig:my_label}, the dark, small regions are circled and overlapped on the SEM image to give the reader a better understanding of what deep dimples are. The larger surrounding region demarkated by the bright lines are the shallow dimples, but in this work we are primarily interested in the deep dimples, hence shallow dimples are not discussed. The annotated SEM images were then cropped into slices of 128x128px size to generate around 17786 images. This methodology of data preparation is popular in medical imaging domain involving whole-scale images (WSI) and the domain of satellite images. For the purpose of training our method, we used around 70\% the total images for training our model, 20\% for validating our model and 10\% of the remaining dataset was reserved for testing our model. Since there is no overlap in the training, validation and testing datasets, we believe the model is able to achieve generalization.  

\begin{figure}
    \centering
\includegraphics[width=.24\textwidth]{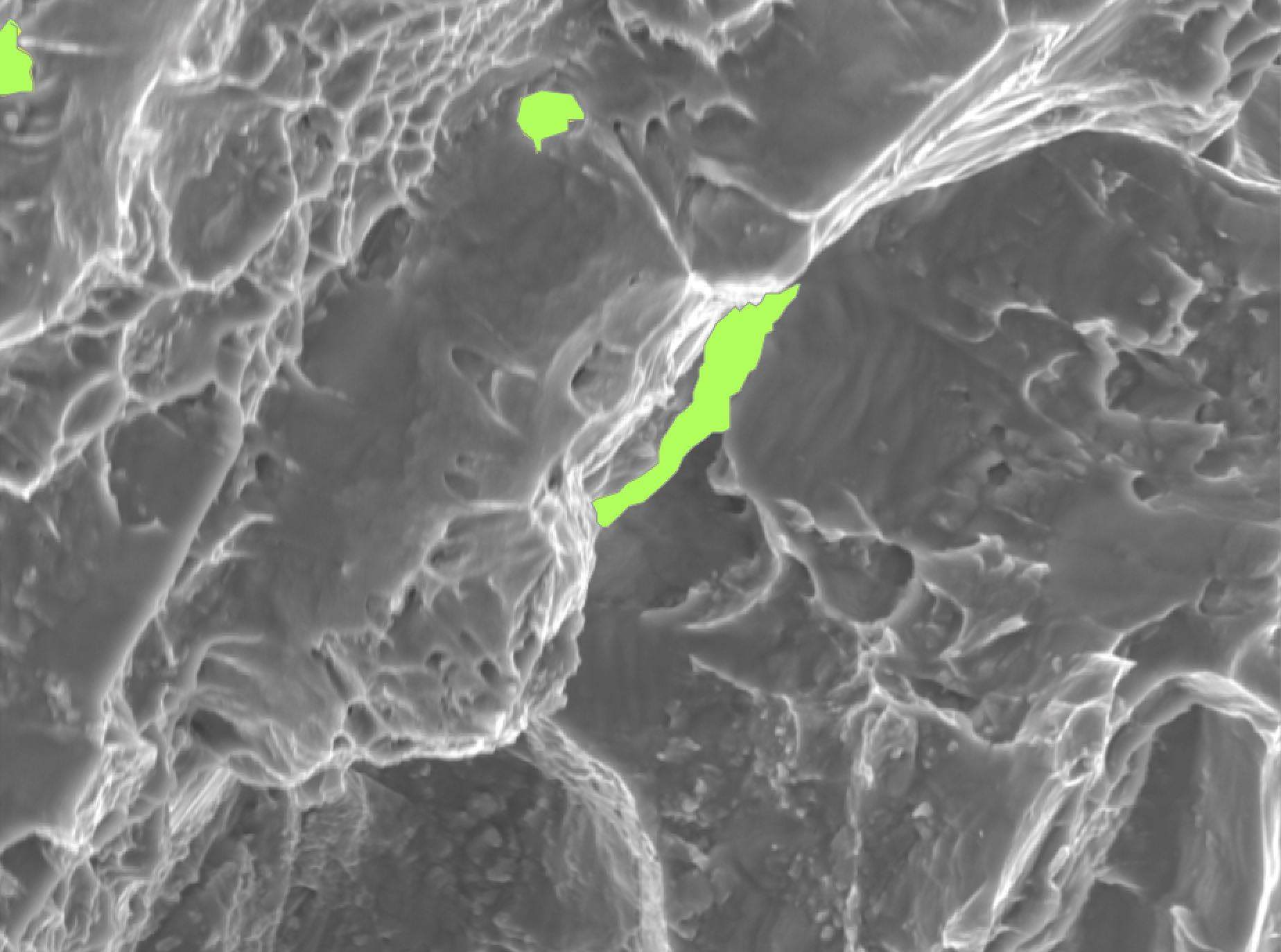}\hfill
    \includegraphics[width=.24\textwidth]{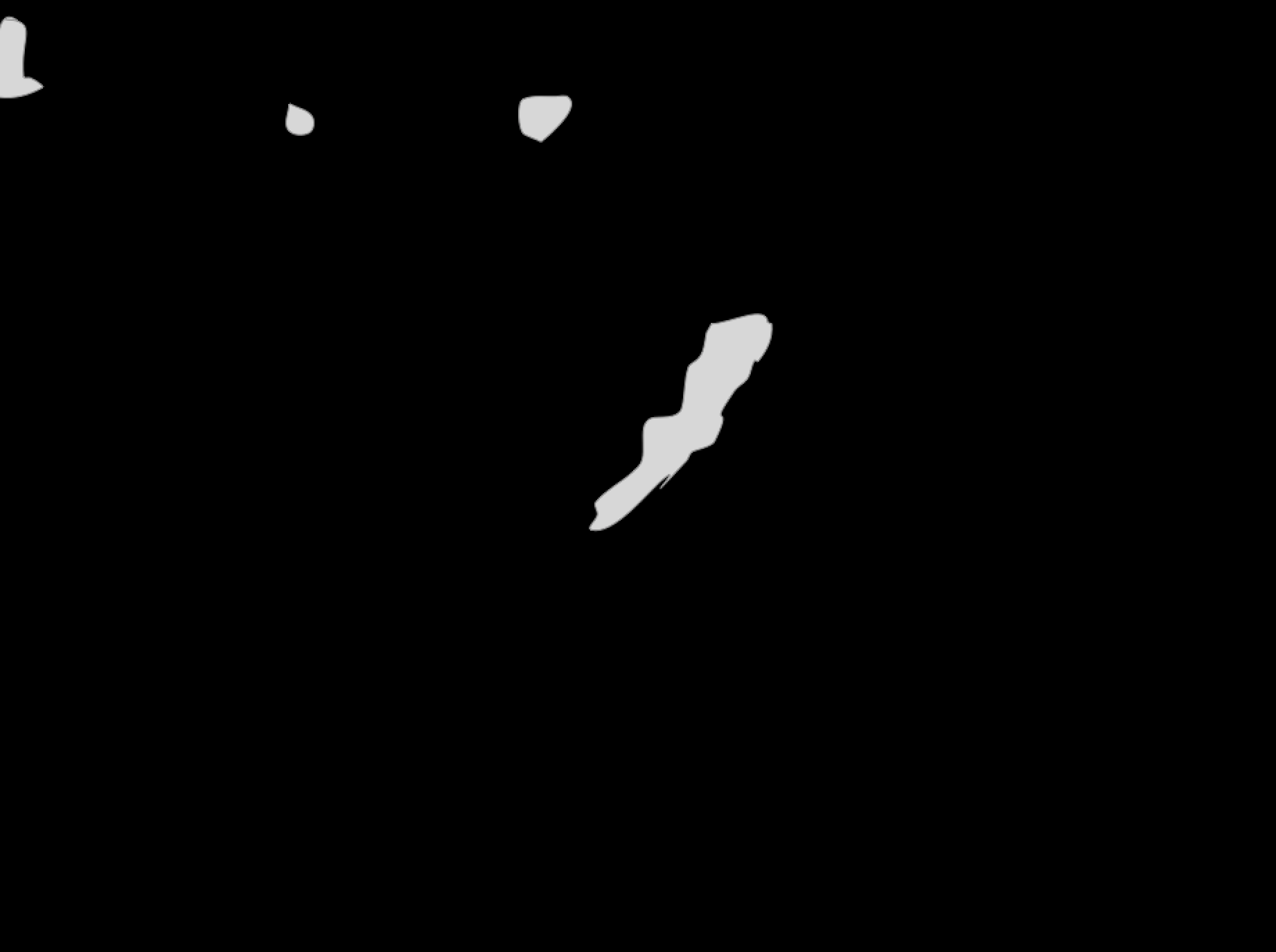}\hfill
    \includegraphics[width=.24\textwidth]{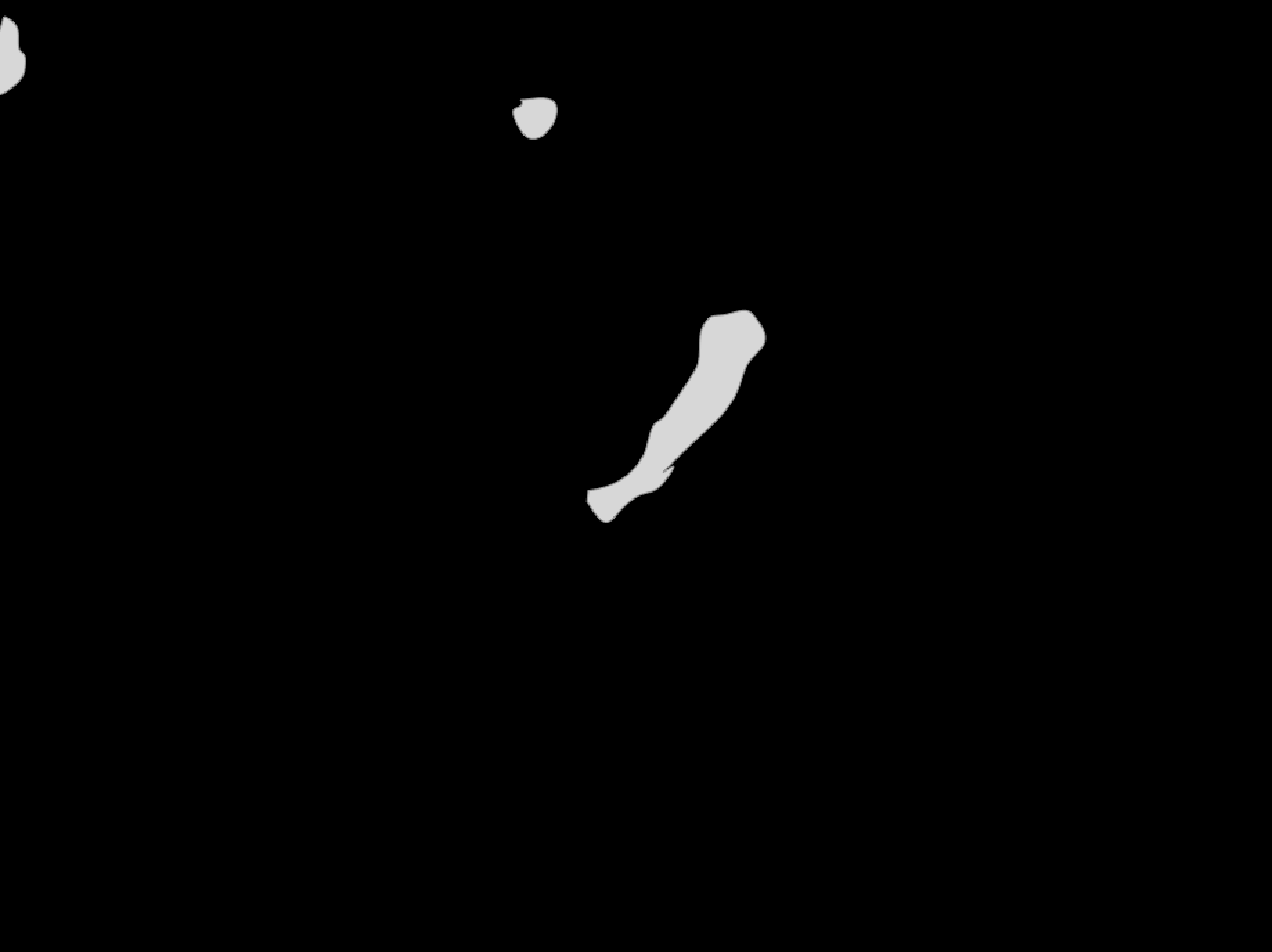}\hfill
    \includegraphics[width=.24\textwidth]{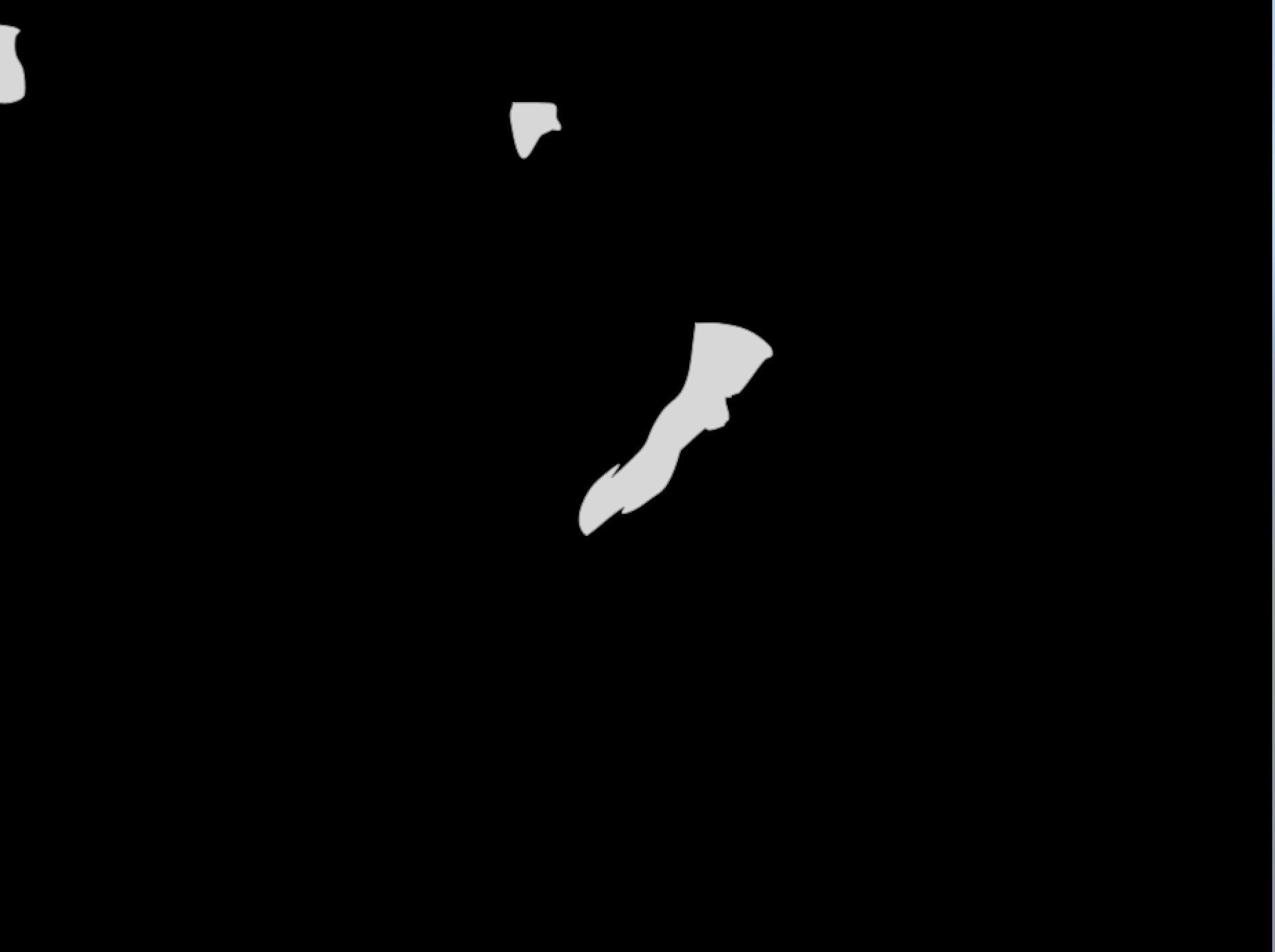}\hfill
    \includegraphics[width=.24\textwidth]{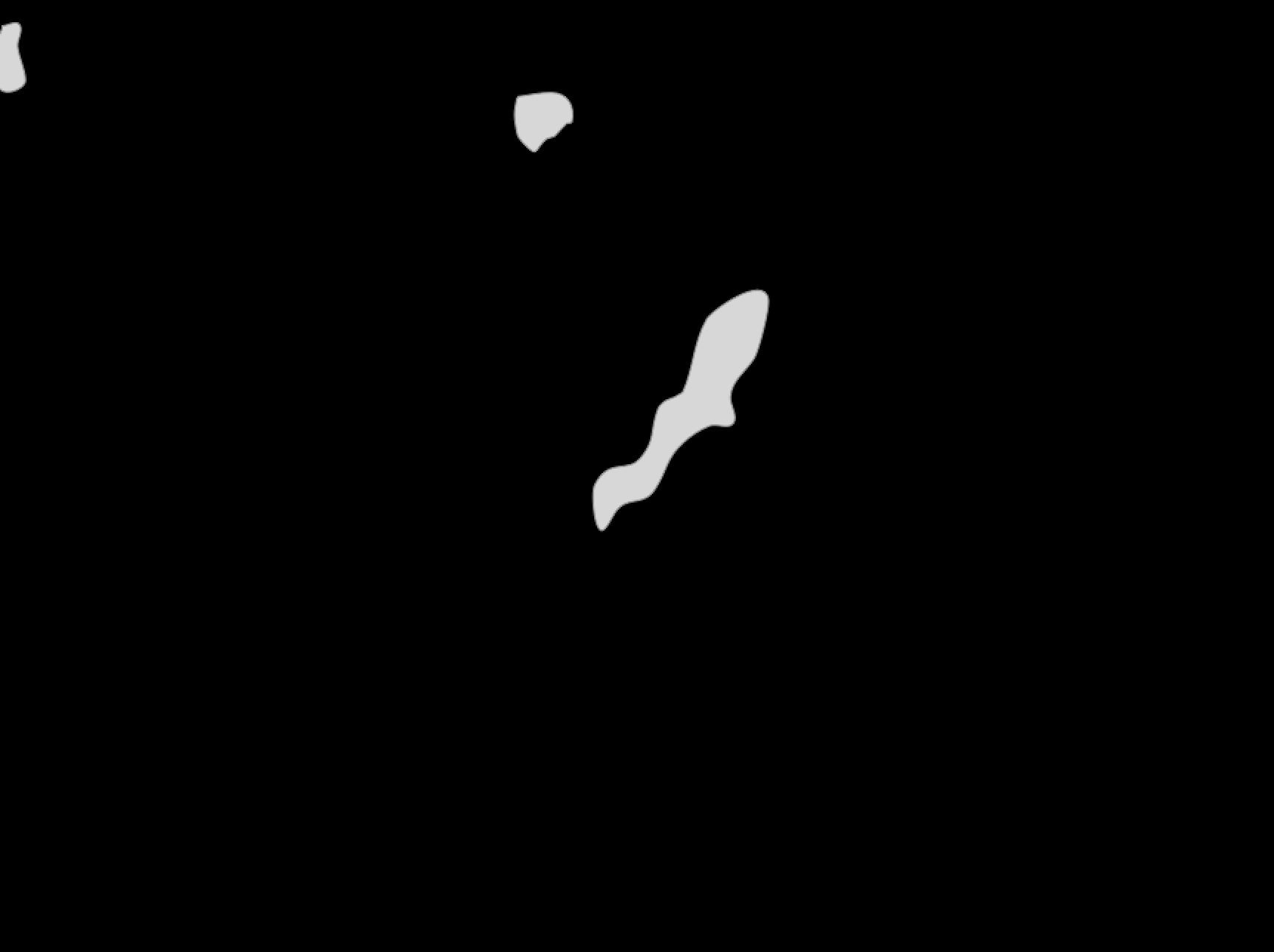}
    \caption{Qualitative Results. From Top to Bottom: GT, U-Net, UNet++, Attention Unet, Res-Unet with Dual Attention \textbf{(ours)}}
    \label{fig:results}
\end{figure}

\subsection{Experiments}

In the experiments, we observed that training the model for a total 150 epochs was enough to reach convergence. The input dimensions of the image was a single channel 128x128 SEM and the output dimension was also a single channel 128x128 segmented mask. After extensive hyper-parameter tuning with optimizer, learning rates and weight decay, we found out that Adam \cite{kingma2014adam} optimizer as an optimizer performs the best with a learning rate of $1e-4$,  a weight decay of $1e-6$, $\beta_1= {0.9} $ and $\beta_2= {0.999}$. We use an exponential decay for learning rate after every 10 epochs. We used a weighted loss function consisting of dice score and binary cross-entropy loss, the weights were generated empirically with $1.25$ for dice-score and $0.95$ for binary cross-entropy loss. The batch size for all the experiments were kept to $1$ due to memory constraints of our GPU. The hyper-parameters were consistent across all the reported methods. We evaluate our model on the basis of dice score \cite{dice1945measures} and compare it to other state-of-the-art methods on semantic segmentation tasks.

\[
L_{total}= \lambda_1 * (1 - L_{dice}) + \lambda_2 * L_{bce}
\newline
\]

The code was written in PyTorch \cite{paszke2019pytorch} and trained on a single Nvidia GeForce GTX 1080Ti. Each experiment took around 6-7 hours.

\[
Dice Score (DSC) = \frac{2 * |A \cap B|}{|A| + |B|}
\]
\begin{center}
     \textit{{where A and B are predicted segmentation map and ground truth, respectively}}
\end{center}

\subsection{Results }

\begin{table}[!h]
\begin{center}
\begin{tabular}{|l|c|}
\hline
Method & Dice Score (DSC) \\
\hline\hline
U-Net & 0.68509  $(\pm 7.88 \%)$  \\
U-Net++ & 0.73423  $(\pm 6.51 \%)$ \\
Attentive U-Net & 0.81630  $(\pm 5.27 \%)$ \\
\textbf{ResU-Net with Dual Attention (ours)} & \textbf{0.86305} $(\pm 5.05 \%)$ \\ 
\hline
\end{tabular}
\end{center}
\caption{Quantitative Results on Microstructures of Ti alloys}
\label{results2}
\end{table}

\begin{figure}
    \begin{center}
        \includegraphics[width=.24\textwidth]{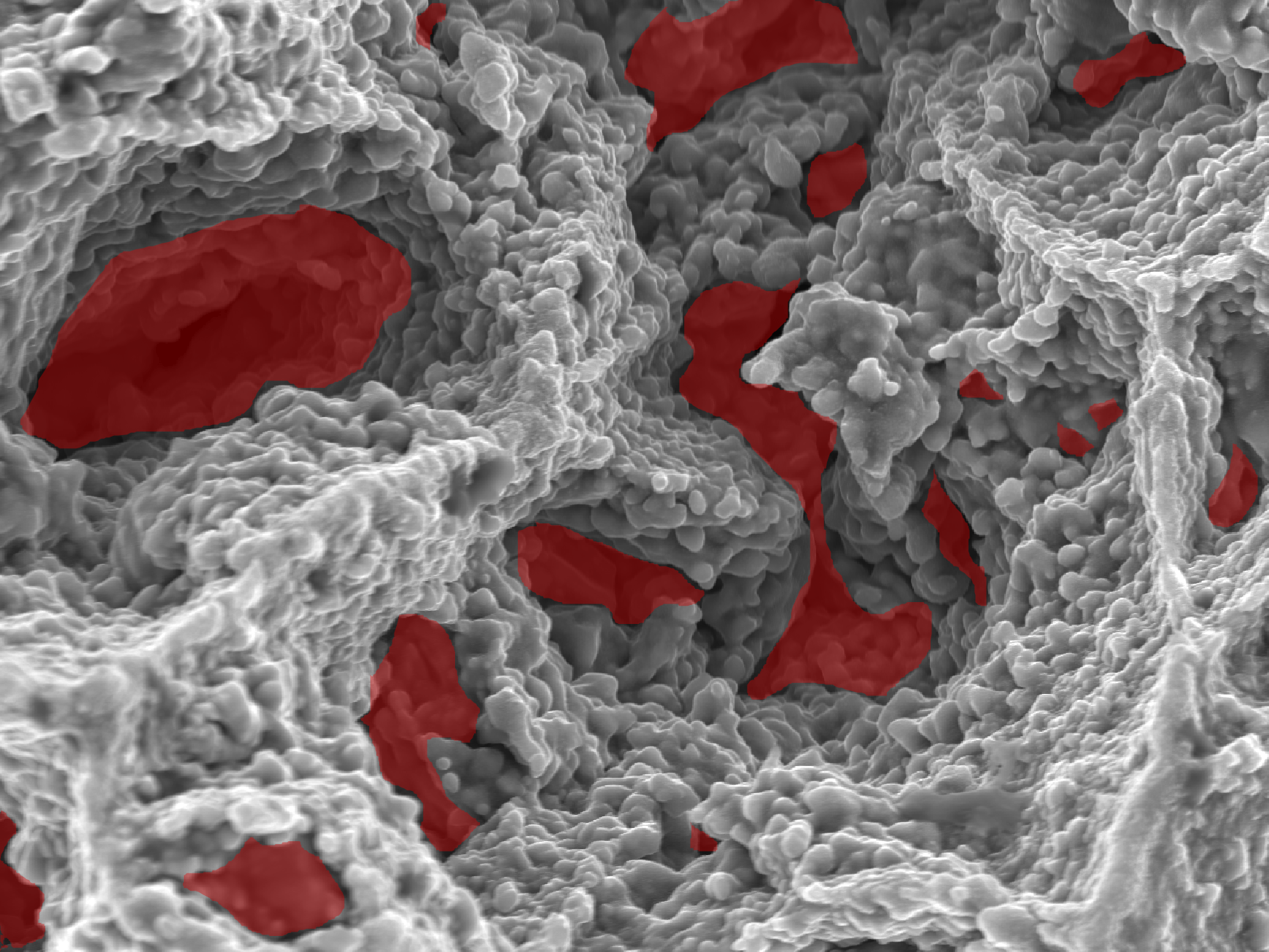}\hfill
    \includegraphics[width=.24\textwidth]{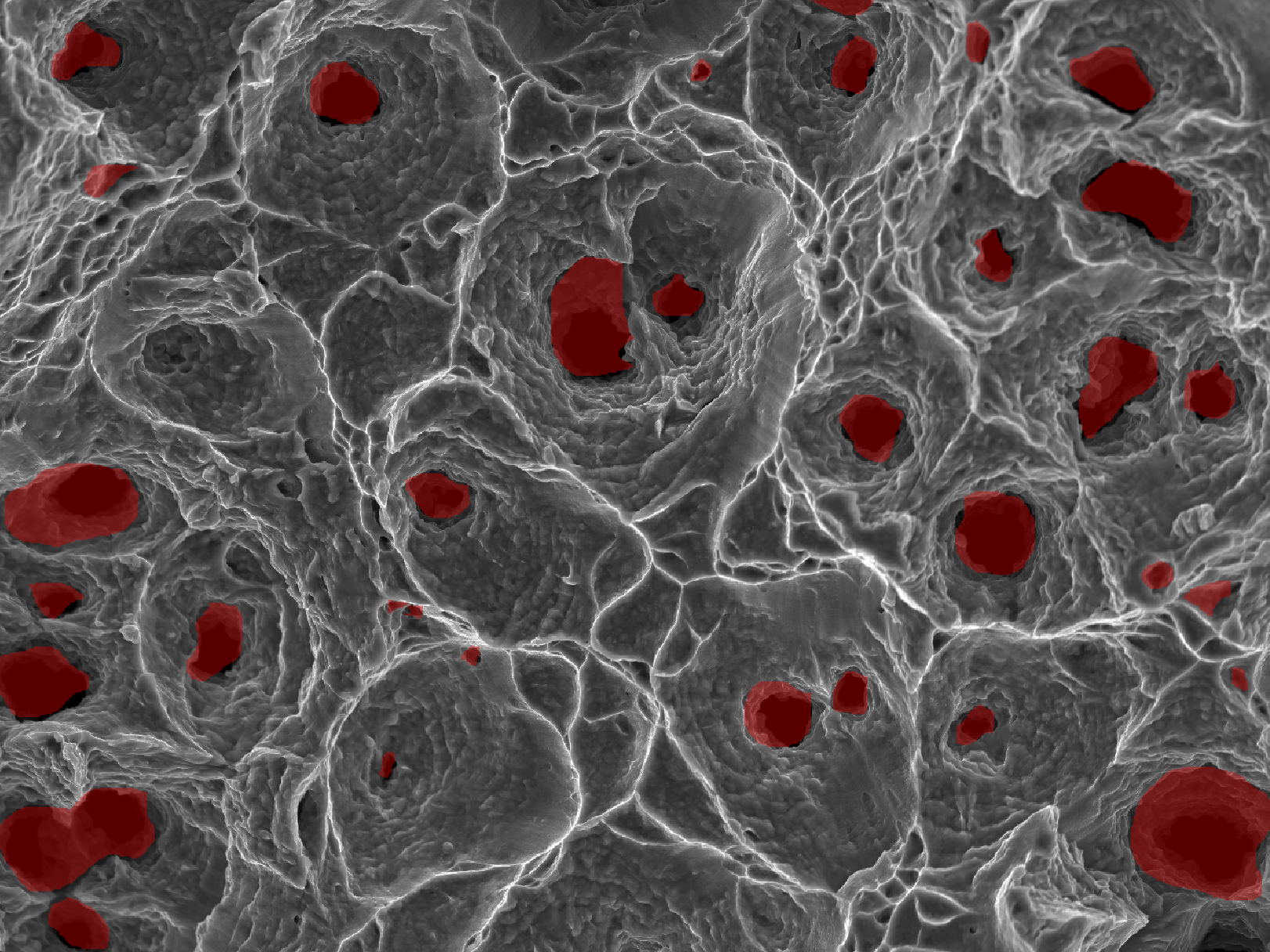}\hfill
    \includegraphics[width=.24\textwidth]{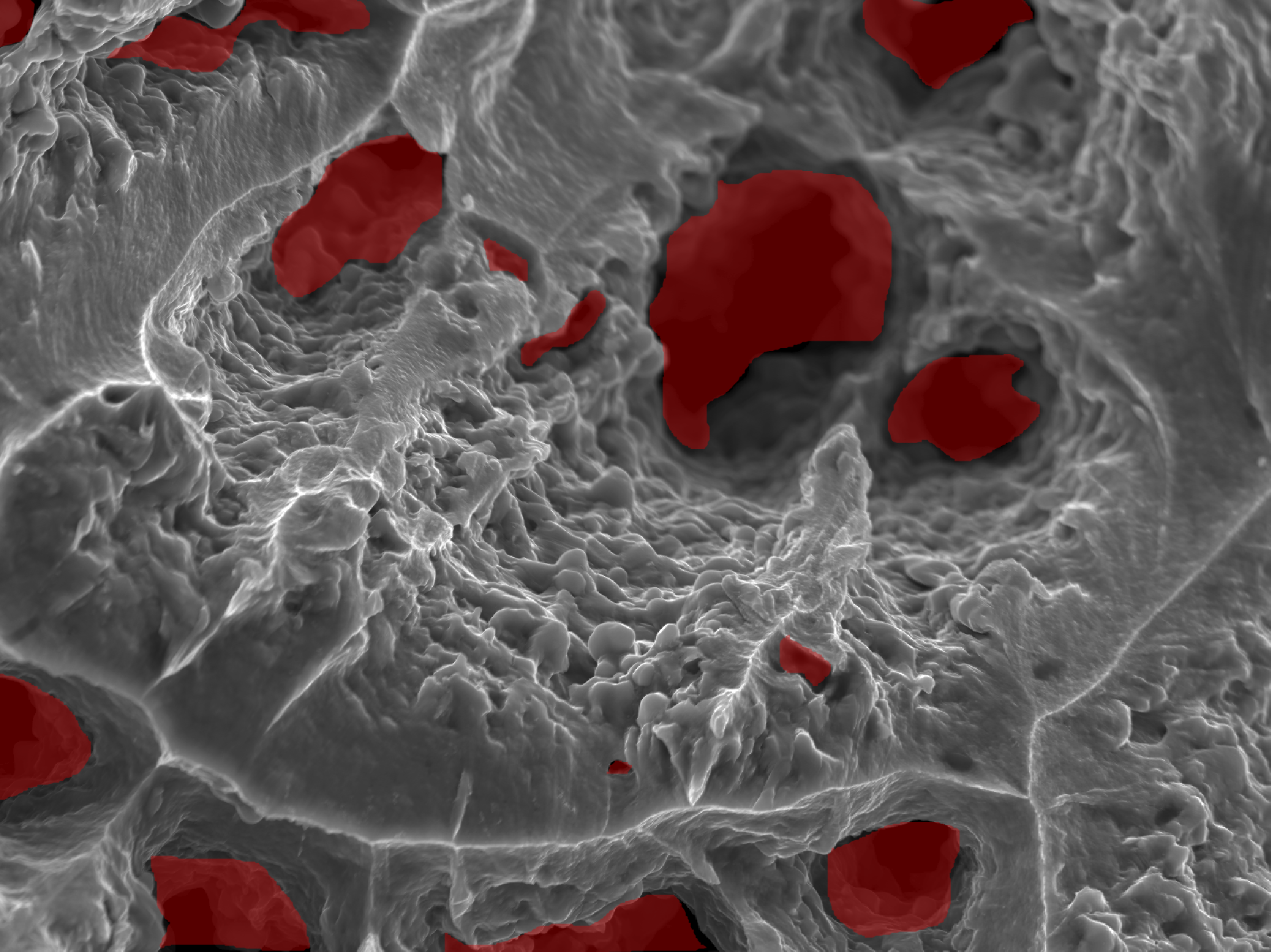}\hfill
    \includegraphics[width=.24\textwidth]{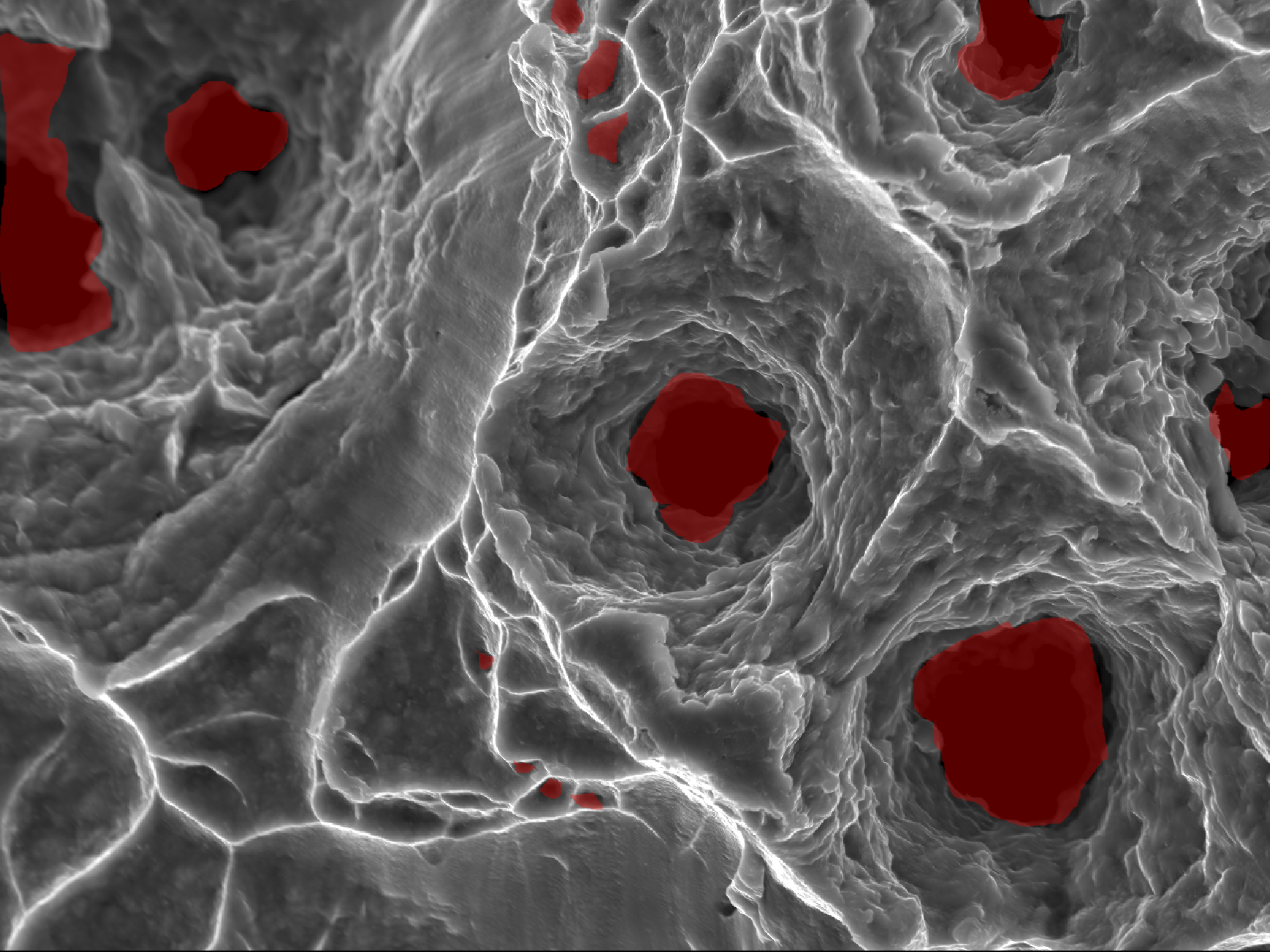}\hfill
    \includegraphics[width=.24\textwidth]{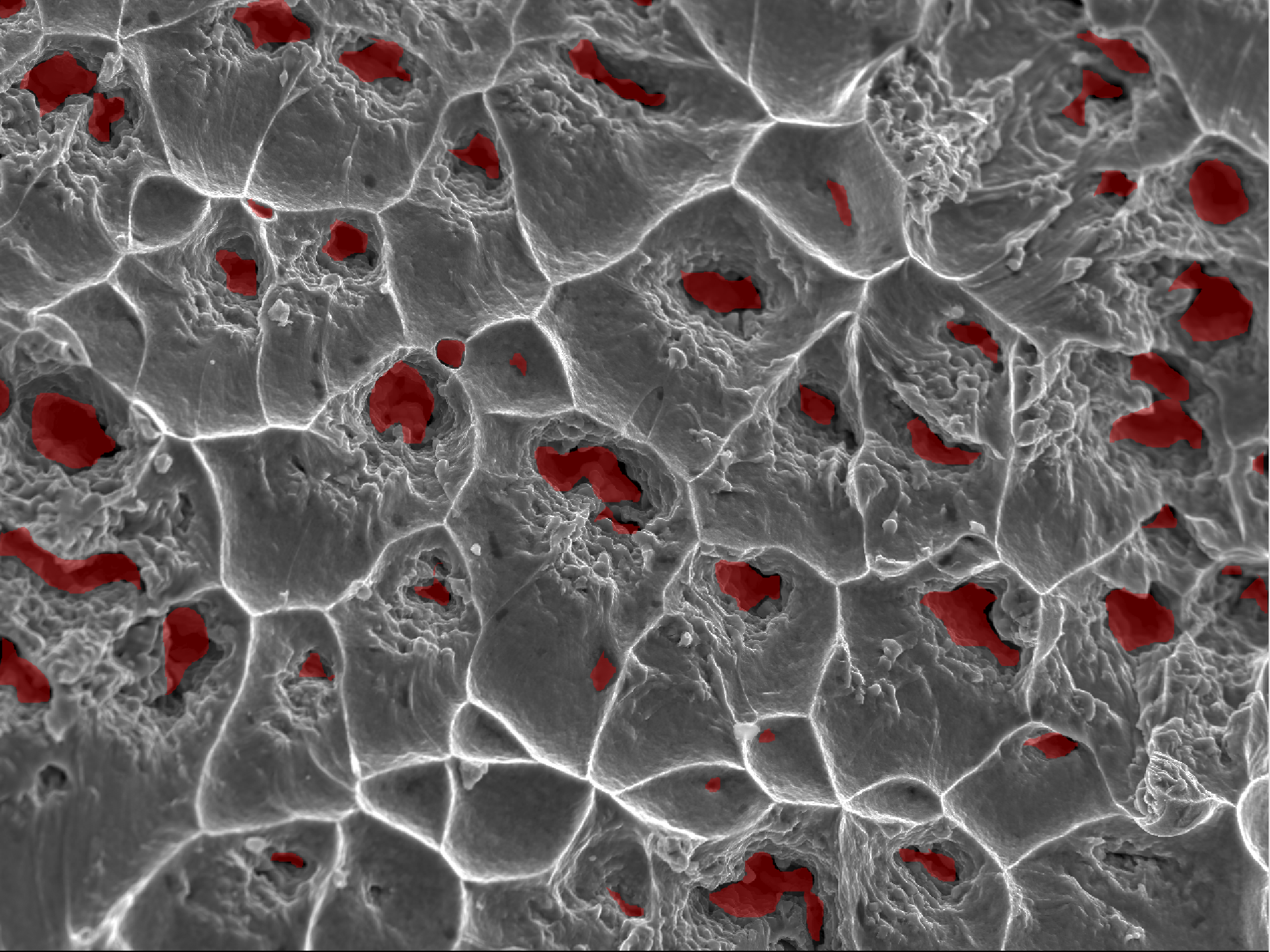}\hfill
    \includegraphics[width=.24\textwidth]{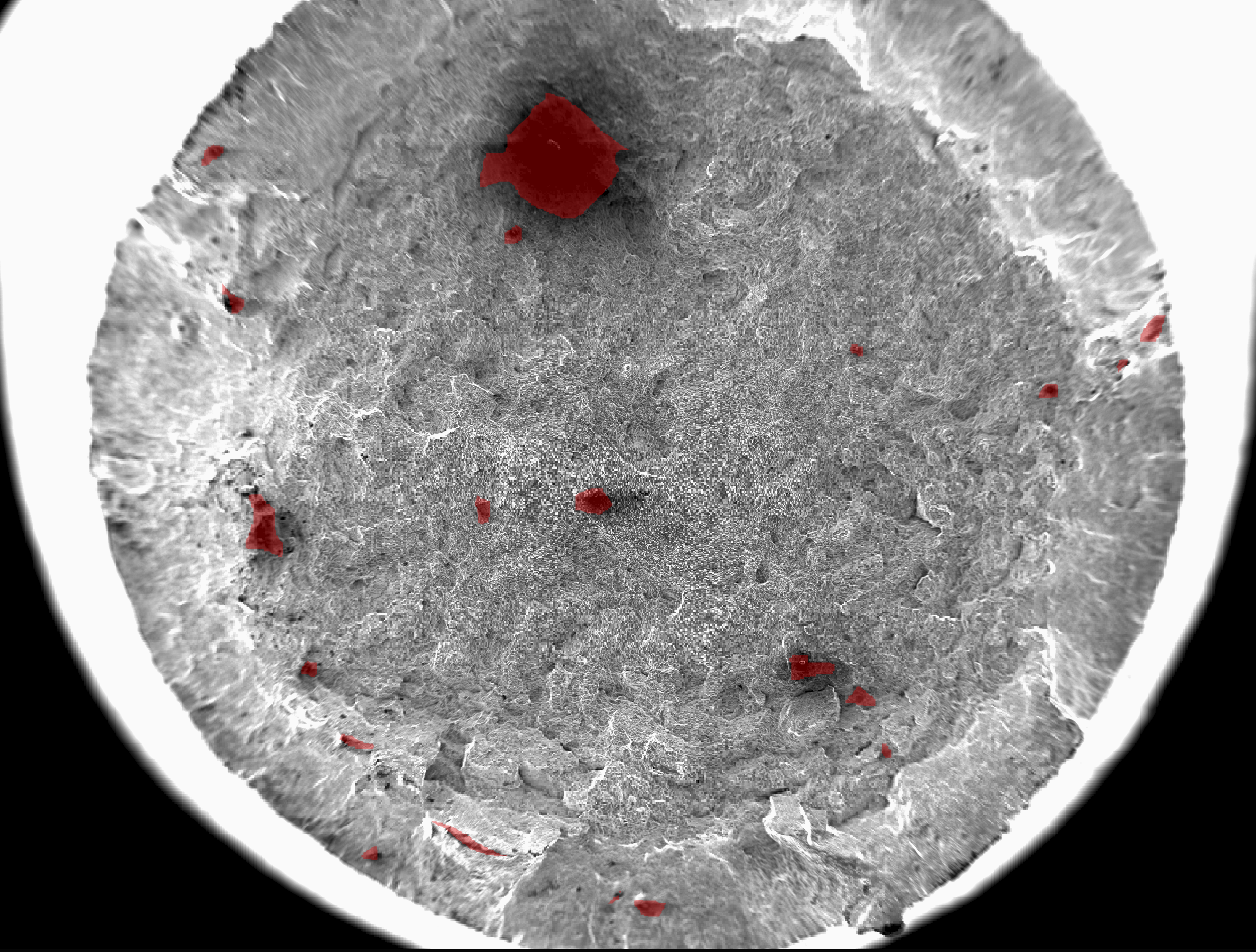}\hfill
    \includegraphics[width=.24\textwidth]{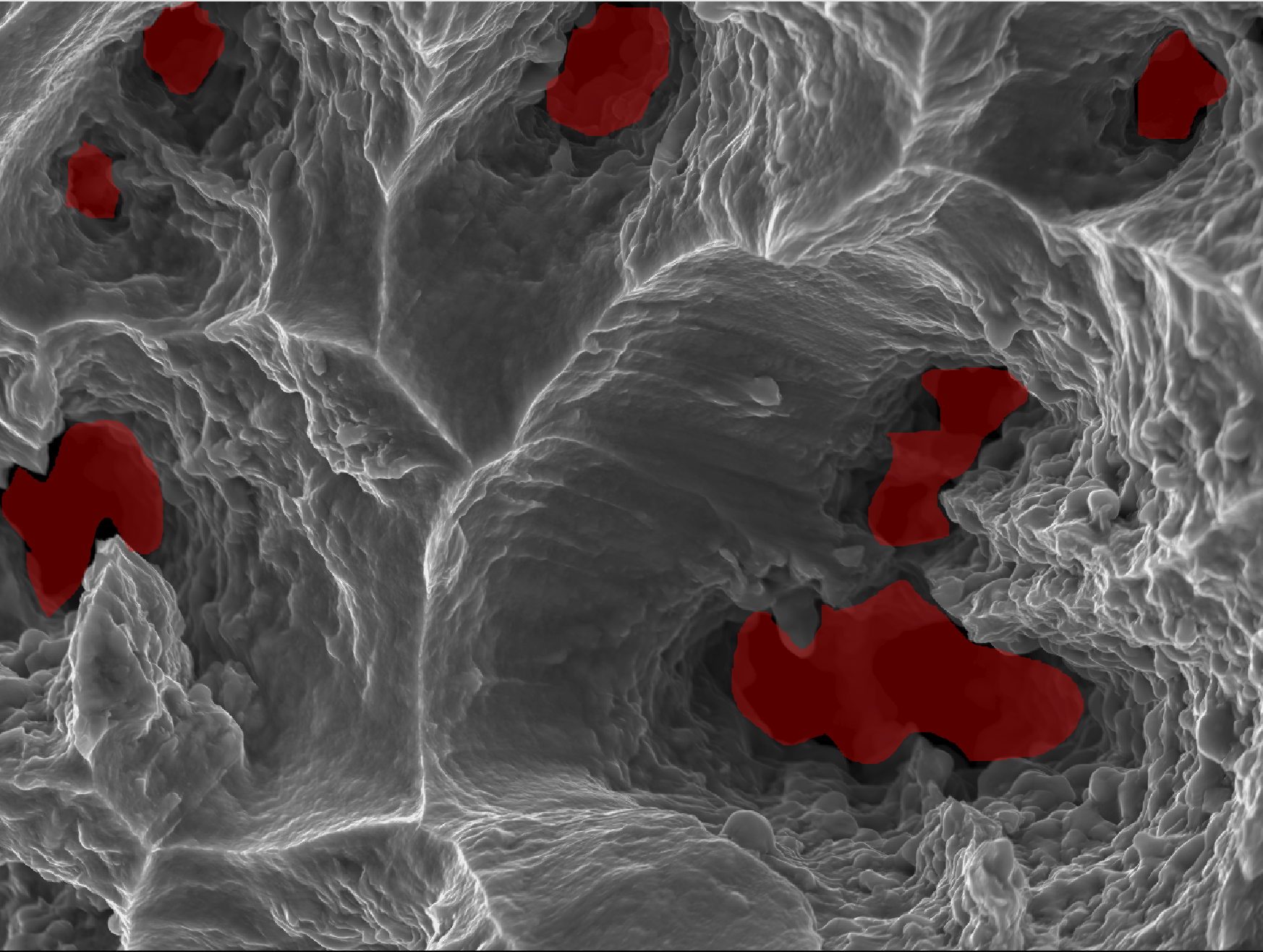}\hfill
    \end{center}

    \caption{Predictions of our model on SEM images.}
    \label{fig:results1}
\end{figure}

We evaluate the performance of our proposed model and the baseline model on the widely used metric of Dice-coefficient. The results of various methods are tabulated in \hyperref[results2]{table 1}. It's visible from the quantitative and qualitative results \ref{fig:results}, \ref{fig:results1} that our proposed model performs the best as compared to other previous established methods. After getting the segmentation results, the results can be analyzed to understand the materials inherent microstructure, grain boundary strength as well as the conditions leading to the failure of the material. 

Quantifying the probability of occurrence of the mechanisms of ductile fracture is cumbersome and highly biased on the user, this makes the task at hand more challenging. Moreover, the size of the features on the fracture surface changes drastically from one material to the other. Since, simple hand-crafted features based model or deep learning based classification or object detection algorithms  are not able to effectively tackle these challenges, we use semantic segmentation algorithms to classify every pixel in the SEM images which allows for the topographic characterization of the fracture surface, making this approach best suited for a fractographic analysis. This method tries to learn to classify the background pixels even though they do not follow a certain pattern, which leads to certain misclassifications. The results presented herein can be improved, given a larger, annotated, training set. Depending on the area occupied by the deep dimples or shallow dimples (beyond the scope of this work) present in the SEM image, we can analyze if the material underwent brittle or ductile fracture and what mechanism did the material follow. Thus, this work aims to reduce the time and effort of material science researchers for fractographic analysis.

\section{Conclusion}

In this work, we have elucidated how fractures occur in the material and how they can be fatal while under service. We have presented an overview of the traditional as well as modern approaches involved in detecting of these defects on the microscopic scale. We also present new methods for the segmentation of deep dimples, which may serve as the first step towards categorizing the type of defect a material had by analyzing the segmented fractographic surface of the SEM image. This work presently focuses on dimple detection for ductile materials like Ti alloys, but can be easily extended to other kinds of defects and other materials too. This is a robust method and the results can be improved provided the availability of large data. This work is an aim to foster machine learning and deep learning principles in automating the traditional methods applied in material science for fractography or material forensics. We hope to see better models in the future which can obtain the desired results with limited data and can lead to reduced human labour and material wastage, increased efficiency during production of new materials.

{\small
\bibliographystyle{unsrt}
\bibliography{egpaper}
}

\end{document}